\documentclass[twoside,reqno]{HERON}

\usepackage{epsfig,cite,colordvi}
\usepackage{graphicx}
\usepackage{url}
\usepackage{amssymb,amsmath,amscd,epsf}
\usepackage{times}
\usepackage{makeidx}
\begin{document}

\title{Pairing Theory of the Wigner Cusp}

\runningheads{Pairing Theory of the Wigner Cusp}{K. Neerg\aa rd}

\begin{start}

\author{K. Neerg\aa rd}{1}

\index{Neerg\aa rd, K.}

\address{Fjordtoften 17, 4700 N\ae stved, Denmark}{1}

\begin{Abstract}

Subtracting the Coulomb energy from the mass of a nucleus results in
what may be called the Coulomb reduced mass. In 1936, Bethe and Bacher
suggested that the latter increases from $N=Z$ approximately
quadratically in $N-Z$, where $N$ and $Z$ are the numbers of neutrons
and protons. Myers and Swiatecki found in 1966 a marked deviation from
this rule; for small $|N-Z|$ the Coulomb reduced mass rises more
rapidly. They called the apparent extra binding energy in the vicinity
of $N=Z$ the Wigner energy. It will be shown that this nonanalytic
behaviour of the mass as a function of $N-Z$, referred to as the
Wigner cusp, arises naturally when the pairing force is treated in the
random phase approximation (RPA). In the limit of equidistant single
nucleon levels the increment of the Coulomb reduced mass from $N=Z$ is
approximately proportional to $T(T+1)$, where $T$ is the isospin,
equal to $|N-Z|/2$ in the ground state of a doubly even nucleus. This
provides a microscopic foundation of taking the macroscopic symmetry
energy to have this form.

Excitation energies proportional to $T(T+1)$ resemble the spectrum of
a quantal, axially symmetric rotor. In 1999, Frauendorf and Scheikh
identified the superfluid pair gap as the deformation which gives rise
to an analogous rotation in isospace. Recent work by Bentley and
Frauendorf in collaboration with the speaker applies a Strutinskij
renormalised independent nucleons plus pairing Hamiltonian to the
description of nuclei in the the vicinity of $N=Z$. The theory
includes a Strutinskij renormalisation of the RPA contribution to the
total energy. This theory reproduces quite well the empirical masses
in the vicinity of $N=Z$ for $A \ge 24$, including the Wigner cusps
and the splitting of the lowest levels with $T=0$ and 1 in the doubly
odd $N=Z$ nuclei. While it is crucial for this result that the liquid
drop symmetry energy is similar to the symmetry part of the
Strutinskij counterterm to the RPA energy in being proportional to
$T(T+1)$ rather than $T^2$, large shell corrections modify this bulk
behaviour. The RPA correction makes a contribution of about 1~MeV to
the $T = 0$ doubly odd doubly even mass differences. For the relative
masses of doubly even nuclei it is insignificant.

\end{Abstract}

\end{start}

\section{Introduction}

In 1966, analysing the mass data of the time, Myers and Swiatecki
discovered a ``sharp trough along $N = Z$ occurring in the masses of
the lighter nuclei'' remaining after ``the experimental masses in the
range $A = 4$ to $A = 58$ were corrected for all known effects
(liquid-drop binding and shell effects deduced from nuclei with
$N = Z$)''~\cite{ref:My66}. It is important for the following to
notice that the symmetry energy term in their ``liquid-drop binding''
is proportional to \linebreak $(N-Z)^2$. Referring to
Ref.~\cite{ref:Wi37}, they name the extra term they include in their
mass formula to account for this anomaly the ``Wigner term'', and it
is customary to refer accordingly to the trough itself as the ``Wigner
cusp''.

The reference to Wigner points to a result in Ref.~\cite{ref:Wi37}
based on the supposition that the two-nucleon interaction is invariant
under arbitrary unitary transformation of the four-dimensional space
of the nucleonic spin and isospin. Under this and some other
simplifying assumptions Wigner derives that the isospin dependent part
of the total interaction energy of a nucleus is proportional to
$T(T+4)$, where $T$ is the isospin, in the ground state of a doubly
even nucleus equal to $|N-Z|/2$. It is known by now that the
two-nucleon interaction does not have this simple structure. Redoing
Wigner's derivation under the assumption of only isobaric invariance,
I show in an appendix to Ref.~\cite{ref:Ne09} that in this case a
factor $T(T+1)$ replaces Wigner's $T(T+4)$. In both cases the estimate
applies only to the \emph{interaction} part of the total energy.
Wigner estimates the contribution from the nucleonic \emph{kinetic}
energy by the Thomas-Fermi model. This gives a $T$-dependent term
proportional to $T^2$ so that if the interaction energy is proporional
to $T(T+1)$ then the $T$-dependent part of the \emph{total} energy is
proportional to $T(T+X)$ with $X < 1$. Empirically, the kinetic and
interaction terms make about equal contributions to the symmetry
energy~\cite{ref:Bo69}.

The nuclear physics literature since 1966 has many attemps of
explaining the Wigner cusp. My allotted time does not allow me to
discuss it all; I must refer you to an extensive review in
Ref.~\cite{ref:Ne09}. Presented here is my own take on the issue.

\section{Superfluid isorotation}

The empirical evidence points to something like a $T(T+1)$ law for the
$T$-dependent part of the Coulomb reduced mass of a nucleus. Thus, for
example, the semiempirical mass formula of Duflo and
Zuker~\cite{ref:Du95} has symmetry terms of this form. Frauendorf and
Sheikh noticed that this is similar to the spectrum of an axially
symmetric rotor and identified the nuclear superfluidity as the
deformation in isospace that could give rise to an analogous
isorotation~\cite{ref:Fr99,ref:Fr00}. In fact, in a product of neutron
and proton Bardeen-Cooper-Schrieffer (BCS) states the expectation
value of the pair field isovector
\begin{equation}\label{eq:vecP}
  \vec P = - 2^{\frac32} i \sum_{p<q}
    \langle p | t_y \vec t\, | \bar q \rangle a_p a_q
\end{equation}
is perpendicular to the expectation value of the isospin $\vec T$. The
pair field therefore precesses about $\vec T$ in a way analogous to
the precesssion of the single-nucleon potential well of the axially
deformed nucleus about the angular momentum. In Eq.~\eqref{eq:vecP}
the vector $\vec t\,$ is the single-nucleon isospin, $a_p$ annihilates
a nucleon in the member $|p\rangle$ of an orthonormal basis of
single-nucleon states and the bar denotes time reversal.

A microscopic version of this picture is explored in
Refs.~\cite{ref:Ne02,ref:Ne03,ref:Ne09,ref:Ne16}. I consider a
Hamiltonian
\begin{equation}\label{eq:H}
  H = \sum_j h_j - G \, \vec P^\dagger \! \cdot \! \vec P
      + \kappa \sum_{j<k} \vec t_j \cdot \vec t_k ,
\end{equation}
where $j$ and $k$ label the nucleons and $h$ is a single-nucleon
Hamiltonian commuting with $\vec t$. By adding a constant, replacing
$\sum_j h_j$, $\vec P$ and $\vec T = \sum_j \vec t_j$ by their
expectation values in a Bogolyubov quasinucleon vacuum and
constraining the expectation values of $N$ and $Z$ by Lagrangian
multipliers, one arrives at the Routhian~\cite{ref:Ne75}
\begin{equation}\label{eq:R}
  R = \left\langle \sum_j h_j \right\rangle - G |\langle \vec P \rangle|^2
      + \tfrac12 \kappa |\langle \vec T \rangle|^2
      - \lambda_n \langle N \rangle - \lambda_p \langle P \rangle . 
\end{equation}
Minimisation of $R$ gives essentially the BCS theory. Zero point
oscillation about the minimal $R$ is taken into account in the random
phase approximation (RPA). This gives for the total energy the
expression
\begin{equation}\label{eq:E}
  E = \sum v^2 \epsilon - \frac{|\vec\Delta|^2}G + \tfrac12 \kappa T^2
      + c + \tfrac12 \left( \sum \omega - \sum \omega_0 \right) ,
\end{equation}
where $\epsilon$ is an eigenvalue of $h$ counted with multiplicity,
$v^2$ is the BCS occupancy, possibly different for neutrons and
protons, $\vec \Delta = G \langle \vec P \rangle$, $T = |N-Z|/2$, the
contant $c$, which does not depend on $T$, accounts for exchange
contributions to $\langle H \rangle$ including a compensation for the
constant added in the derivation of Eq.~\eqref{eq:R}, $\omega$ denotes
an RPA frequency and $\omega_0$ is a two-quasinucleon energy. The
expectation values are taken in the BCS state. The BCS equations
separate into independent ones for neutrons and protons and the RPA
equations into independent ones for two-quasineutron, two-quasiproton
and quasineutron-quasiproton excitations.

If $\Delta_n = - \, 2^{-\frac12} G \langle P_- \rangle \ne 0$, a
Nambu-Goldstone mode~\cite{ref:Na60,ref:Go61} arises from the
invariance of $R$ under transformations $\exp i \alpha N$, and
similarly if \linebreak
$\Delta_p = 2^{-\frac12} G \langle P_+ \rangle \ne 0$. If $T > 0$ or
$T = 0$ and $\Delta_n = \Delta_p \ne 0$, there is also a
``quasi''-Nambu-Goldstone mode with frequency
$|\lambda_n - \lambda_p|$ following from the relation
\begin{equation}
  [ -\lambda_n N - \lambda_p Z , T_- ]
    =  ( \lambda_n - \lambda_p ) T_- .
\end{equation}

The last term in Eq.~\eqref{eq:H}, which I call the
\emph{symmetry force}, turns out to have no other effect than adding
its eigenvalue $\frac12 \kappa [ T(T+1) - 3A/4 ]$ to the total energy.
It may thus be included a posteriori. For the moment I set $\kappa =
0$ so that my Hamiltonian is the bare, isobarically invariant pairing
Hamiltonian. The RPA theory of this Hamiltonian was first discussed by
Ginocchio and Wesener~\cite{ref:Gi68}.

\section{Uniform single-nucleon spectrum}\label{sec:uniform}

Important qualitative insight is gained from the case of a uniform
single-nucleon spectrum with constant level density $g$. This is
studied in Ref.~\cite{ref:Ne16}. Presently, let it be assumed for
simplicity that the Kramers degenerate single-nucleon levels are
$\epsilon = 1/(2g), 3/(2g), 5/(2g), \dots, (2\Omega_\tau - 1)/(2g)$,
where $\tau = n, p, np$ refer to the neutron and proton BCS + RPA and
neutron-proton RPA calculations, respectively, and $\Omega_n = N$,
$\Omega_p = Z$, $\Omega_{np} = A/2$. Displacing the single-nucleon
spectrum adds $A$ times the displacement to the total energy. It is
understood that $N$ and $Z$ are even.

The result of replacing summation over $\epsilon$ by integration is
\begin{equation}\label{eq:uniform}
  E = \sum_{\tau=n,p} \left(
        g \int_0^{\frac{\Omega_\tau}{2g}} \epsilon \, d \epsilon
      + E_\text{BCS,$\tau$} \right)
      + \sum_{\tau=n,p,np} E_\text{RPA,$\tau$} ,
\end{equation}
where $E_\text{BCS,$\tau$}$ and $E_\text{RPA,$\tau$}$ are given by
closed, analytic expressions. For $\tau = n$ and $p$ these expressions
are functions of $G$, $g$ and $\Omega_\tau$ while $E_\text{RPA,$np$}$
also depends on $\delta\lambda_{np} = \lambda_n - \lambda_p$. By
expressing $E$ as a function of $T$ for a constant $A$ one gets
\begin{gather}
  \sum_{\tau=n,p} \left(
        g \int_0^{\frac{\Omega_\tau}{2g}} \epsilon \, d \epsilon
      + E_\text{BCS,$\tau$} \right)
    \approx \frac { \displaystyle \frac {A^2} 4 + T^2 } {2g} , \\
  E_\text{RPA,$n$} + E_\text{RPA,$p$} + E_\text{RPA,$np,T=0$}
    = \text{constant} , \\ \label{eq:T/2g}
   E_\text{RPA,$np$} -  E_\text{RPA,$np,T=0$}
    \approx \frac T {2g} .
\end{gather}

The contribution~\eqref{eq:T/2g} comes from the quasi-Nambu-Goldstone
mode. It is seen to add a term to the ``kinetic'' symmetry energy
$T^2/(2g)$ which renders also this part of the total symmetry energy
proprotional to $T(T+1)$. Eq.~\eqref{eq:T/2g} is a special case of a
general relation
\begin{equation}
  \omega_\text{quasi-NG} = \frac d {dT} E_\text{mean field} .
\end{equation}
Hence if $E_\text{mean field} - E_\text{mean field,$T=0$}
\propto T^2$ ---which holds necessarily when the mean field state is
not an eigenstate of $T_z$ and $T$ is sufficiently small---then
$E_\text{mean field} - E_\text{mean field,$T=0$}
+ \frac12 \omega_\text{quasi-NG}\propto T(T+1)$~\cite{ref:Ne03}.
Marshalek made the analogous observation for spatial
rotation~\cite{ref:Ma77}. As the RPA energy atop a mean field
theory has always the form of the last term in Eq.~\eqref{eq:E}, it
follows that if the mean field energy rises quadratically in $T$ and
the remainder of the RPA energy is constant then the $T(T+1)$
propotionality of the symmetry energy is exact in the mean field plus
RPA approximation. In fact this remainder is, in the uniform case,
\emph{not} a constant but adds a negative term, which is proportional
to $T^2$ to the lowest order of $T$. This correction is largest in the
lightest nuclei. With realistic parameters its absolute value amounts
to at most 3 MeV for $T \le 0.2 A, A \ge 24$.

\section{Woods-Saxon levels, $\ \kappa > 0$}\label{sec:WS}

Calculations with a Woods-Saxon single-nucleon spectrum and $\kappa
>0$ are reported in Ref.~\cite{ref:Ne09}. I refer to Figs.~3 and~4 of
that article. When the single-nucleon spectrum is derived from a
deformed Woods-Saxon potential ($A = 48, 68, 80$) the expectations
from the uniform case are roughly borne out as seen, for example, from
$\frac12 \omega_\text{quasi-NG} / (E(T) - E(0)) \approx 1/3 =
T/(T(T+1))$ for $T = 2$. But when the $N = Z$ nucleus is doubly magic
($A = 56, 100$) an entirely different picture emerges. Then the
\emph{kinetic} symmetry energy is almost linear in $T$. Quadratic
terms arise from the symmetry force and to some extent from the onset
of BCS pairing when the neutron and proton Fermi levels move from the
macic gap into the shells above and below. This gives rise to a large
$X$ in an approximation of the total symmetry energy by an expression
proportional to $T(T+X)$.

The reason for the linearity in $T$ is that the neutron excess is
generated by a promotion of nucleons accros the magic gap in the
single-nucleon spectrum. Each proton transformed into a neutron adds
to the total independent-nucleon energy an amount approximately equal
to the width of the gap.

\section{Accuracy of the BCS + RPA, BCS crtiticality}

It is well known that the BCS gap parameter $\Delta_n$ vanishes for
coupling constants $G$ below a critical value $G_{\text{crit},n}$ of
the order of the spacing of the single-nucleon levels surrounding the
neutron Fermi level, and similarly for protons. (It follows that
$G_{\text{crit},\tau} = 0$ if the Fermi levels is within a spherical
subshell and also in the uniform approximation.) The accuracy of the
BCS + RPA can be tested by comparison with a numeric minimisation of
the Hamiltonian. Bentley and Frauendorf made such minimisations in
valence spaces of six and seven Kramers and isospin degererate
single-nucleon levels~\cite{ref:Be13}, and comparisons with the BCS +
RPA were done by Bentley~\etal~\cite{ref:Be14}; see Fig.~1 of the
latter article for at fairly extreme case.

It turns out that globally, the BCS + RPA reproduces the exact energy
very well. It can be shown to be asymptotically exact for $G \to
\infty$~\cite{ref:Gi68,ref:Be14}. However, a plot of $E$ as a function
of $G$ has sharp dips at the critical $G$ while the exact curve goes
smoothly through these points. This gives rise to fairly large local
deviations. As a remedy Bentley~\etal\ introduce an interpolation of
the last term in Eq.~\eqref{eq:E}. In the calculations reported below
interpolation is applied for $0.5 \, G_{\text{crit},\tau} < G < 2 \,
G_{\text{crit},\tau}$ when $\tau = n$ or $p$, and in the union of these
intervals when $T = 0$ and $\tau = np$.

The singularity of the RPA energy at $G = G_{\text{crit},\tau}$ is
well known from the literature~\cite{ref:Ba70} and has a natural
explanation in terms of the emergence of Nambu-Goldstone modes at
these point~\cite{ref:Be14}.

\section{Strutinskij renormalisation}

Bentley~\etal\ apply a Strutinskij renormalisation to
the theory presented so far~\cite{ref:Be14}. My present
rendering of the method and its results includes some as yet
unpublished updates.

The total energy is written
\begin{equation}\label{eq:renorm}
  E = E_\text{mic} - \tilde E_\text{mic} + E_\text{LD} .
\end{equation}
where $E_\text{mic}$ and $\tilde E_\text{mic}$ are given by
Eqs.~\eqref{eq:E} and~\eqref{eq:uniform}, respectively, except that the
first term in the bracket in Eq.~\eqref{eq:uniform} is replaced by
\begin{equation}
  \int_{-\infty}^{\tilde\lambda_\tau} \tilde g_\tau(\epsilon) \epsilon \, d
\epsilon .
\end{equation}
Here $\tilde g_\tau(\epsilon)$ and $\tilde \lambda_\tau$ are
Strutinskij's smooth level density and chemical
potential~\cite{ref:St67}. For clarity I write in the present context
$\tilde E_\text{BCS,$\tau$}$ and $\tilde E_\text{RPA,$\tau$}$ for the
terms $E_\text{BCS,$\tau$}$ and $E_\text{RPA,$\tau$}$ in
Eq.~\eqref{eq:uniform}. In these terms I use 
$g = \tilde g_\tau(\tilde\lambda_\tau)$ with $\tilde g_{np}(\epsilon)$
and $\tilde \lambda_{np}$ to be defined below. In Ref.~\cite{ref:Be14}
a heuristically based approximation of $\tilde E_\text{RPA,$np$}$ is
employed.

The liquid drop energy $E_\text{LD}$ is taken in the
form~\cite{ref:Du95}
\begin{equation}\label{eq:LD}\begin{split}
   E_\text{LD} =\
     & - \left(a_v - a_{vt} \frac{T(T+1)}{A^2} \right) A \\
     & + \left(a_s - a_{st} \frac{T(T+1)}{A^2} \right) A^{2/3} B_s
       + a_c \frac{Z(Z-1)}{A^{1/3}} B_c ,
\end{split}\end{equation}
where $B_s$ and $B_c$ are the usual functions of
deformation~\cite{ref:Bo39}. A conventional Nilsson-Strutinskij
calculation~\cite{ref:Be10} supplies the deformations and
single-nucleon spectra, and the different spectra for neutrons and
protons are used everywhere except in the calculation of
$\tilde E_\text{RPA,$np$}$, where average neutron and proton levels
are employed. The calculations reported in Ref.~\cite{ref:Be14} use
average levels throughout. The variable $\delta\lambda_{np}$ in
$\tilde E_\text{RPA,$np$}$ (see the text after Eq.~\eqref{eq:uniform})
is set to $\tilde\lambda_n - \tilde\lambda_p$ with $\tilde\lambda_n$
and $\tilde\lambda_p$ calculated from $\tilde g_{np}(\epsilon)$ while
$\tilde\lambda_{np}$ corresponds to filling $A/2$ nucleons into the
smooth average spectrum. In all parts of $E_\text{mic}$ the $A/2$
lowest levels of either kind of nucleon are included in the
calculation, and I take accordingly $\Omega_\tau = A/2$ for all $\tau$
in the calculation of $\tilde E_\text{mic}$.

In the calculation of $E_\text{mic}$ for odd $N = Z$ and $T = 0$ one
neutron and one proton contribute the Fermi energy and the BCS and
RPA schemes are applied to the remaining nucleons populating the
remaining levels. The lowest state with $T = 1$ in such nuclei is
assumed to differ in energy from its isobaric analogue with $Z -1$
protons only by the difference in liquid drop Coulomb energy.

\section{Data and parameters}

I denote by $E(A,T)$ the ground state energy for
\mbox{$N + Z = A, (N - Z)/2 = T$}, where $N$ and $Z$ are even, while
$E^\ast(A,T)$ is the lowest energy for odd \linebreak $N = Z = A/2$
and isospin $T$. The following combinations of these energies are
considered.
\begin{itemize}
\item $\Delta_\text{oo-ee}
  = E^\ast(A,0) - \tfrac12 [E(A-2,0) + E(A+2,0)]$ .
\item $E^\ast(A,1) - E^\ast(A,0)$ .
\item The constants $\theta$ and $X$ defined by
$$\begin{gathered}
  E(A,T) = E_0 + \frac{T(T+X)}{2\theta}
           + a_c \frac{Z(Z-1)}{A^{1/3}} B_c , \\
  T = \begin{cases}
        0,2,4, & A \equiv 0 \mod 4, \\
    1,3,5, & A \equiv 2 \mod 4.
    \end{cases}
\end{gathered}$$
\end{itemize}
The five parameters in Eq.~\eqref{eq:LD} were fitted to the the values
of $E(A,T)$ for\linebreak $24 \le A \le 100, 0 \le N - Z \le 10$ that
were measured according to the 2012 Atomic Mass Evaluation
(AME12)~\cite{ref:Au12}. The resulting rms deviation is 0.875 MeV,
which is somewhat better than 0.950 MeV achieved in the calculations
of Ref.~\cite{ref:Be14}. The parameters $G_1$ and $\chi$ in
$G = G_1 A^\chi$ were fitted to the empirical values of
$\Delta_\text{oo-ee}$ and $E^\ast(A,1) - E^\ast(A,0)$ for
$26 \le A \le 98$ in so far as they can be derived from data from
AME12 and the National Nuclear Data Center~\cite{ref:NNDC}. Here the
resulting rms deviation is 0.708 MeV. Because the two sets of
parameters are not independent, the fits were repeated alternately
until both converged. The complete set of resulting parameters is
$a_v,a_{vt},a_s,a_{st},a_c,G_1,\chi =
15.07,107.4,16.04,133.6,0.6506,7.232,-0.7604$,
all of them except $\chi$ in MeV.

\section{Results}

\begin{figure}
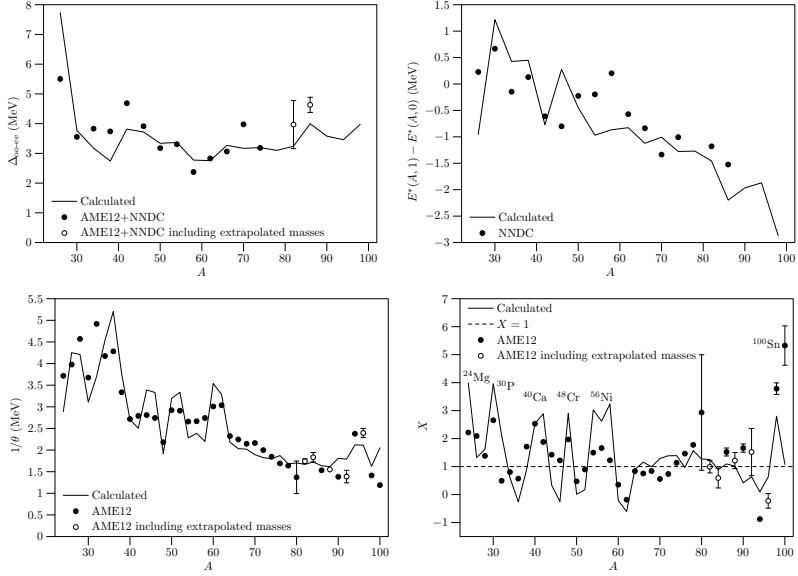
\centering
  \includegraphics[height=.2\textheight]{oo-eebw}\quad
  \includegraphics[height=.2\textheight]{1-0bw}\\[1ex]
  \includegraphics[height=.2\textheight]{rthbw}\quad
  \includegraphics[height=.2\textheight]{Xbw}
  \caption{\label{fig:comp}Four differential mass combinations as
     functions of $A$.}
\end{figure}

Figure~\ref{fig:comp} shows comparisons of the calculated and measured
values of $\Delta_\text{oo-ee}$, $E^\ast(A,1) - E^\ast(A,0)$,
$1/\theta$ and $X$. While no perfect agreement is achieved, the model
evidently accounts well for qualitative aspects of the variations with
$A$. In the calculations, these aspects have simple explanations in
terms of the shell structure. I discuss in particular the peaks in the
plot of $X$; see the figure.
\begin{itemize}
\item $A = 40, 56, 100$: The $T = 0$ nucleus is doubly magic. The
mechanism that gives rise to a large $X$ is discussed in
Sec.~\ref{sec:WS}.
\item $A = 30$: Here the mechanism is similar. For $T = 1, 3, 5$ a
pair of neutrons occupy the $2s_{1/2}$ spherical orbit and isospin
beyond $T = 1$ is generated by a promotion of nucleons from the
$1d_{5/2}$ to the $1d_{3/2}$ shell.
\item $A = 24, 48$: Deformation is important. The $T = 0$ nucleus has
a large deformation while for $T = 2$ and 4 the nucleus is almost or
completely spherical. The cost in deformation energy required to make
$T = 2$ and 4 gives the large $X$.
\end{itemize}

Given these microscopic explanations of some large $X$ one might ask
whether the $T(T+1)$ proportionality of the liquid drop symmetry
energy is required. The answer is affirmative as seen from what
happens if it replaced by a $T^2$ proportionality and the liquid drop
parameters are refitted: (1)~The rms deviation of the doubly even
masses increases from 0.875 MeV to 1.073 MeV.
(2)~$E^\ast(A,1) - E^\ast(A,0)$ decreases by 1--2~MeV thus becoming
negative also in the lightest nuclei. (3)~$X$ decreases by 0--2 units
rendering the calculation mostly below the data.

The marked underestimate of the $X$ measured for $A \approx 100$
might reflect an inaccurate representation by the Nilsson model of the
shell gaps at $N = Z = 50$.

\section{Role of the RPA correction}

\begin{figure}\centering
  \includegraphics[width=.7\textwidth]{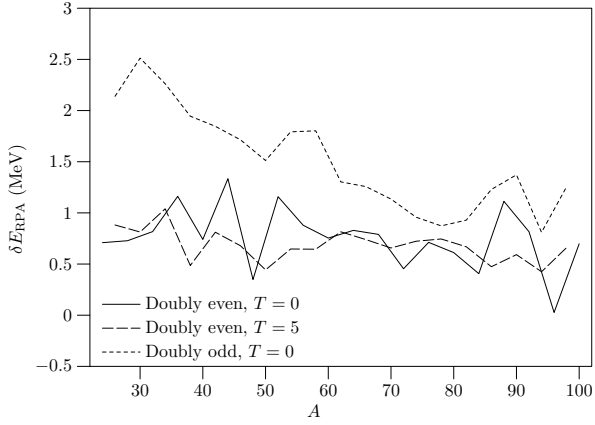}
  \caption{\label{fig:dRPAbw}The RPA correction in several cases.}
\end{figure}

The RPA correction $\delta E_\text{RPA} =
\sum_{\tau=n,p,np} (E_{\text{RPA},\tau} - \tilde E_{\text{RPA},\tau})$,
where\linebreak $\sum_{\tau=n,p,np} E_{\text{RPA},\tau}$ is the last
term in Eq.~\eqref{eq:E}, is plotted in Fig.~\ref{fig:dRPAbw} in
several cases. It is seen that in the doubly even nuclei it is largely
constant with an average of about 0.7~MeV. Therefore differential
quantities like $1/\theta$ and $X$ are mainly unaffected by $\delta
E_\text{RPA}$. The changes in the plots of these quantities when
$\delta E_\text{RPA}$ is set to zero are barely visible. In other
words, \emph{the shape of the Wigner cusp is well reproduced by a
Nilsson-Strutinskij calculation with only a BCS pairing correction}.

In the doubly odd $T = 0$ states $\delta E_\text{RPA}$ is considerably
larger, decreasing from about 2.3~MeV to about 1.2~MeV in the range of
the plot. This influences the fit of the pair coupling constant $G$ to
the measured $\Delta_\text{oo-ee}$. More precisely, \emph{the RPA
correction reduces the required pair coupling constant}.

Both the general positivity of $\delta E_\text{RPA}$ and its larger
value in the doubly odd nuclei can be understood as a result of an
effective dilution of the single-nucleon spectra near the Fermi
levels~\cite{ref:Ne16}. In the doubly even nuclei this dilution stems
from the equilibration of the deformation. In the doubly odd nuclei a
further dilution results from the inaccesibility of the Fermi levels
to the pairing force.

\section{Conclusions}

The insight gathered on this tour may then be summarised as follows.
\begin{itemize}
\item Calculations with an RPA correction added to the BCS pairing
correction conventionally employed in Nilsson-Strutinskij calculations
account well for the variation with $A$ of the pattern of masses near
$N = Z$.
\item The RPA correction is insignificant for reproducing the doubly
even masses and hence for the shape of the Wigner cusp.
\item It is important, however, that the \emph{macroscopic} (liquid
drop) symmetry energy be proportional to $T(T+1)$.
\item This form of the macroscopic symmetry energy is understood
\emph{microscopically}, in terms of the RPA, to result from the
nuclear superfluidity.
\item The variation of the shape of the Wigner cusp is dominated by
shell effects.
\item The RPA correction significally reduces the $T = 0$ binding
in doubly odd nuclei, thus reducing the required pair coupling
constant.
\end{itemize}

It is due to mention that work by Negrea and Sandulescu based on a
picture of quartet condensation addresses the same differential mass
combinations except $E^\ast(A,1) - E^\ast(A,0)$ and reproduces them
equally well~\cite{ref:Ne14}. Noteworthy are the traits both
approaches have in common: Both employ states constructed from
isovector Cooper pairs and both take measures to ensure isobaric
invariance.

\bibliography{rila-paper}

\end{document}